# A Context-Aware and Self-Adaptation Strategy for Cloud Service Selection and Configuration in Run-Time

Asmae Benali[1] & Bouchra El Asri[1]

[1] IMS Team, ADMIR Lab, ENSIAS, Rabat IT Center, Mohammed V University in Rabat, Morocco

Correspondence: Asmae Benali, IMS Team, ADMIR Lab, ENSIAS, Rabat IT Center, Mohammed V University in Rabat, Morocco.

## Abstract

Day after day, the number of mobile applications deployed on cloud computing continues in increasing because of smartphone capabilities improvement. Cloud computing has already succeeded in the web-based application, for that reason, the demand for context-aware services provided by cloud computing increases. To customize a cloud service that takes into account the consumer requirements, which depend on information change, it brings to light many recent challenges to cloud computing about environment-aware, location-aware, time-aware. The cloud provider, moreover, has to manage personalized applications and the constraints of mobile devices in matters of interaction abilities and communication restrictions. This paper proposes a strategy for selecting automatically an appropriate cloud environment that runs out whole requirements, defines a configuration for the associated cloud environment and able to easily adapt to the change of the environment on either the user or the cloud side or both. This process builds on the principles of dynamic software product lines, Agent-oriented software engineering, and the MAPE-k model to select and configure cloud environments according to the consumer needs and the context change.

**Keywords:** autonomic system, cloud computing, context-aware system, dynamic software product line, multi-agent system, variability management

## 1. Introduction

In the cloud-computing model, IT resources are provided as services, which are divided into two service models. A model is a service that is usually denoted by X (XaaS or * aaS), as X represents a layer where there is a service as infrastructure form (IaaS) and another as platform form (PaaS) (Armbrust et al., 2016). At the IaaS level, all the software stack running inside the virtual machine must be configured as well as anything that concerns the infrastructure: number of virtual machines, amount of resources, database configuration, etc. Regarding the platforms provided by PaaS, the configuration concerns only the software that makes up the platform, application ie; application server, database, libraries, etc. Thus, the cloud offers many configuration choices, so that the is deployed and the runtime environments are configurable. As a result of the large gamut of resources at different functionality levels, developers must deal with the variability of the cloud environment at the application deployment. Therefore, Cloud Computing is considered as a runtime adaptive system which can dynamically change their services, regarding user requirements changes. the system and user context are considered as inputs to adjust their behavior to the current status of users (for example, contextual activity). Cloud computing has new challenges when it asks for information about the user's environment in terms of location, device detection, and custom applications that are timed to respond to, for example, mobile device constraints, interaction capabilities, resources and communication constraints.

Selecting an appropriate cloud environment and managing its variability, in the execution phase, induces errors in the configuration phase or complex configuration results which are generally made in an ad hoc manner. Based on the findings of the research areas presented in the precedent section, we have introduced this approach. It was established to automatically select an appropriate cloud environment that fits requirements and to determine a configuration for the cloud environment associated, which can be adapted easily to the change of the environment (client and/or cloud side).

We aim to develop a context-aware process for automated and self-management cloud service selection and configuration based on the principles of dynamic software product lines, agents-oriented software engineering and the MAPE-k model to help its user to personalize the selection and configuration of cloud services according to their needs and their context.

Product line engineering is a popular and widely used technology in the industry (Clements & Northrop, 2002). The fundamental principle of this approach is the reuse of the basic product components and structures and apply planned variabilities to derive various product families. The use of product line engineering approach for product development has attained several benefits in productivity, quality, time, cost (Van der Linden, Schmid, & Rommes, 2007). An essential issue for the success of SPLs is variability management (Clements & Northrop, 2002). It is represented through variation points and variants. In the SPL artifact, a variation point represents a design decision, where each variation point is related to a set of variants, i.e., alternatives. The proposed approach uses dynamic software product lines to assure the reliability of these selection and configuration processes. It leverages also the feature model to present cloud variability and to use environment configuration files.

This framework will be managed autonomously using innovative concepts of self-management.

A Multi-Agent System (MAS) is a set of agents that collaborate and interact with their environment, due to the agent's autonomy, reactivity and mobility (Dinesh Kumar & Ashwin, 2012). It can represent the domain information and to execute necessary action to arrive at particularized goals. Multi-agent technologies provide an effective approach for scalable and open systems that are changed dynamically. Through this work we try to response to the following research questions:

Q1: How can cloud services take the consumer context change into account?

Q2: How can automate dynamically the the variably management of cloud services?

Q3: How can cloud services context-aware at run-time?

The article is structured as follows: Section 2 presents a brief overview of the background used throughout the paper. Motivation and the approach overview are presented in Section 3. Section 4 shows the case study of our approach consisting in deploying a mobile cloud service for Diabetes Management.

In section 5, Related work is discussed. Our conclusions and perspectives are described in Section 6.

## 2. Background and Motivations

In this section, we present briefly DSPLs in literature. Moreover, we describe the basics of the feature model, which is used to model respectively the variability and to generate the reconfiguration plan.

### 2.1 Dynamic Software Product Line

(Clements & Northrop, 2002) defines SPL as "a set of software-intensive systems sharing a common, managed a set of features that satisfy the specific needs of a particular market segment or mission and that are developed from a common set of core assets in a prescribed way."

DSPL is a method that moves the product line engineering process to the runtime phase. It addresses runtime configurations which are directed by variability specifications, ensuring that system adaptations meet the specified requirements. Indeed, a DSPL is a unique system that is talented to adapt its behavior at runtime whereas the SPL engineering process acts only at design time, generating several products of the same family. The SPL and DSPL contain a set of common and managed elements known as features (Clements & Northrop, 2002). The variability model contains a set of variable characteristics that make a product different from the others. Therefore, the SPLs processes generate products by choosing specific values and attributes from the variable features mentioned in the model.

### 2.2 Feature Models

Feature diagrams were introduced firstly in the Feature-Oriented Domain Analysis (FODA) method. (Kang, Cohen, Hess, Novak, & Peterson, 1990) define a Feature Model as "a way to represent the information of all possible configurations for a specific product that can be built. The features are organized hierarchically in a diagram in the form of a tree, where it contains a root element, and any feature can have sub-features, as well as constraints, usually inclusion or exclusion".

Figure 1 exposes an example of a partial feature diagram of an IaaS. A feature diagram is composed of the tree root or concept (IaaS), and its subfeatures showing mandatory, e.g. (OS), optional, e.g. (IDE), and alternative features (Linux, Windows). The relations (edges) between features (nodes) can be: AND, XOR and OPTIONAL relationships.

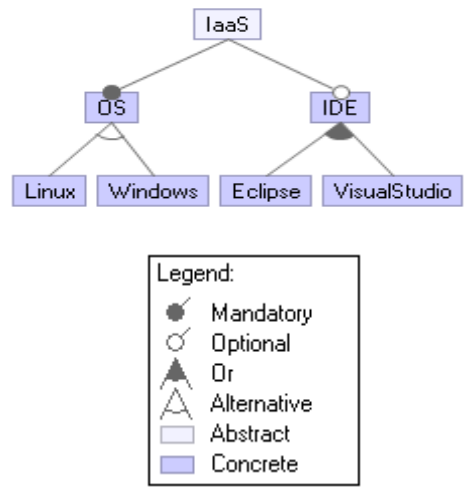

Figure 1. Example of a Feature Model

*2.3 MAPE-K Reference Model*

The MAPE-K reference model (Monitor, Analyze, Plan, Execute and Knowledge) is the most influential feedback control model for self-adaptive systems (Kephart & Chess, 2003).

The Monitor phase permits the collection of information from the software and the runtime environment. It bases on sensors to collect pertinent data. As a result, the plan recognizes a set of modifications to apply to meet the requirements and the manner to make this change. The result of this stage is called an adaptation plan, which uses an auto-configuration and self-optimization technique.

The Analyze phase merges the collected data and the historical data to verify if the requirements are fulfilled or not. In the negative case, it results in a need for adaptation; where the analyze phase demands the Plan phase to build the adaptation plane. This later consists of finding a relevant decision to adapt the software via different algorithms and rules. Ultimately, the Execute phase applies the adaptation plan using effectors to change the concerning element.

### 3. Motivation and Approach Overview

This section presents the motivation for our work. It discusses the challenges to surmount for context-aware cloud service selection and configuration at run-time. Then, an overview of our approach was introduced at the end of this section.

*3.1 Challenges*

The cloud developers have to manage, during the application deployment, a wide range of configurable resources among usable cloud environments. Besides, they must satisfy the customer need and at the same time maximize the profit of the cloud broker. The process goal is to provide a solution to manage the variability of the cloud and user context at the design-time and runtime. Furthermore, the approach has to select and configure autonomously the cloud environment to adapt its context and the context of user change. The following challenges were identified, which the proposed framework must take into consideration, to achieve these objectives:

- ▪ *Challenge 1. Find a suitable environment*

Cloud developers have always to find the cloud providers that affords, despite of the context change in the execution time, all the features required to satisfy the consumer requirements and   to run the user application correctly, like the appropriate database and the virtual machine with at least 3GB of RAM, which is considered as a non-functional requirement.

- ▪ *Challenge 2. Find a valid configuration*

The variability management of the cloud and the user environment is a complex and error-prone task because of the multiple configuration choices which are usually made in an ad hoc manner. Besides that, Developers have not a global view of how a cloud environment is configured. That's why there is a height probability of inconsistencies occurrence between cloud services during the application running. Also, not taking into account the variability of the consumer environment when configuring a cloud service can cause client dissatisfaction and SLA violations. Hence, this second challenge was taken up to find a correct cloud configuration while ensuring the fulfillment of the required features.

- ▪ *Challenge 3. Optimize the configuration*

The customers and providers of the cloud have to establish a Service Level Agreement (SLA) to define the quality of service (QoS). The main goal of cloud providers is the cost minimization and the improvement of the customer satisfaction level (CSL).

- ▪ *Challenge 4. Guarantee the correct configuration*

After finding the optimal configuration that is appropriate for the required functionalities and non-functionality exigences, developers have to avoid errors during the set-up of the configuration files and scripts to make sure to obtain the correct configuration of the cloud environment.

- ▪ *Challenge 5. Keep track of the configuration*

Experimentally, the storage of trace entries on configurations furnishes information for a posteriori analysis. It extends from context conditions to configuration plans.

*3.2 Our Approach*

This section presents our proposed Method as depicted in Figure 2. Our approach is based on the agent-based method to gather information from both the consumer and the cloud environment and also interact for processing tasks. The most systems deployed on cloud computing are autonomous, context-aware, with big data information, interactive and have to provide flexible and customized user services. Agents used in this work precisely to handle the context change at the execution phase and to automatize the selection and configuration process for the cloud providers.

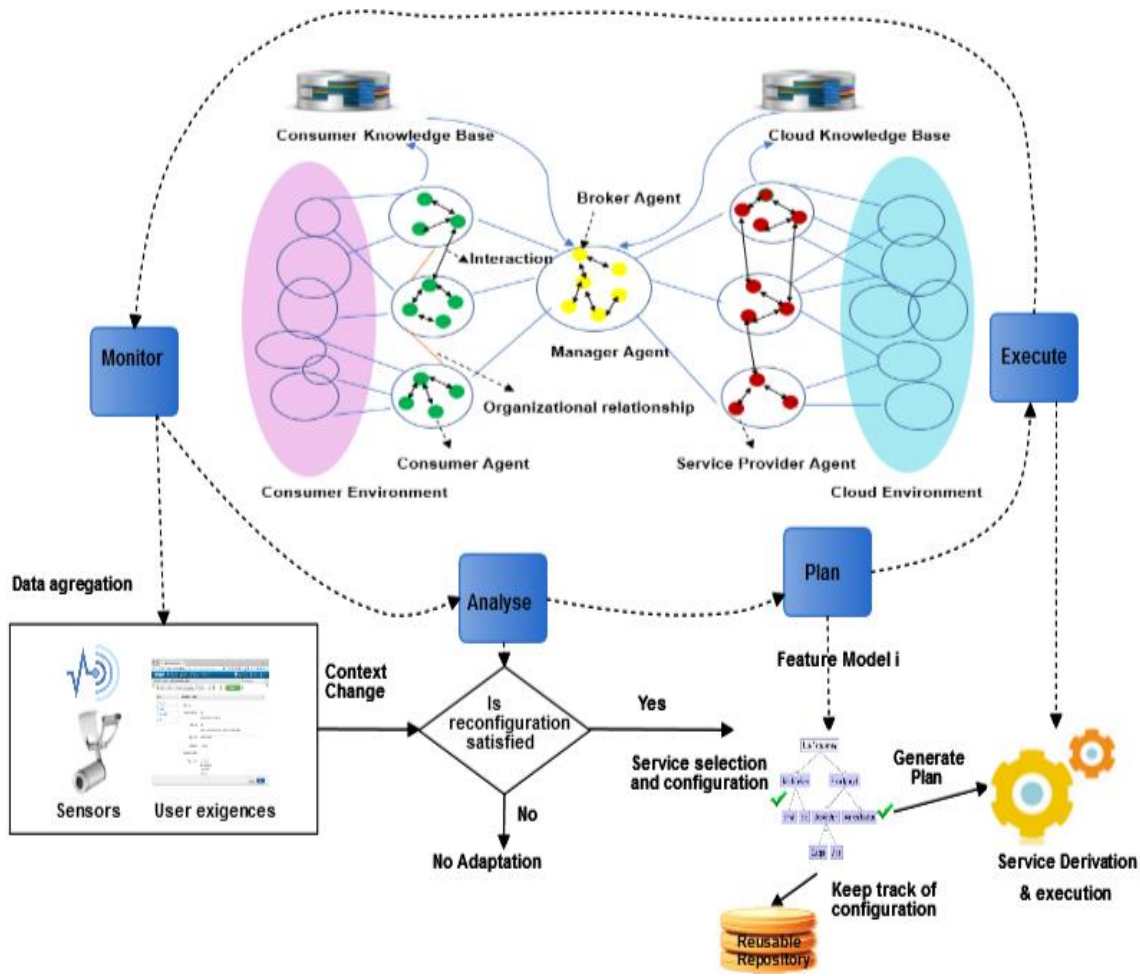

Figure 2. Approach Overview

Besides, we use the feature model approach for variability management by presenting characteristics of the cloud and the consumer environment as features. Moreover, our method provides support for selecting and configuring context-aware cloud services at design and runtime phase by covering the steps of the MAPE-K loop:

***Knowledge.*** The Consumer agent is responsible for interacting and collecting the requirements and the context data of the consumer. It defines the set of characteristics and the cloud service requirements from a graphical user interface. Besides, the consumer agent gathers the information that can be influenced by the continuous change of consumer context using sensors. Afterword, a set of pre-processing data mining techniques are applicated to the data saved in the Consumer knowledge base, such remove unnecessary and duplicate information to reduce the data size before sending it to the cloud and prepare it for the processing step. These data represent dynamic variation points, OCL constraints, software architecture, and resource information. Each Agent in our approach has its proper capabilities Knowledge to cooperate with the other agents and to connect to its environment. The provider service agents can interact through the different cloud layers. In this work, the agents interact between us basing on Jade ACL message Structure.

Table 1. Interaction Message Format of agents

| Message Type | Address Receiver | Address Sender | Content | Conversation ID |
|---|---|---|---|---|

As already indicated that the cloud architecture is composed of three different layers, for that reason each layer has its agents who are different from the agents of other layers. Agents, in the IaaS layer, have the role to provide by the basic information about the cloud resources. In the PaaS layer, agents are responsible for the deployment and execution environments programs that are used to implement the application. Concerning the last layer, SaaS layer, the agents optimize the utilization of the applications mentioned as services and maintain the QoS.

***Monitor.*** The consumer agent, in this phase, takes into charge the providing the agent manager by consumer context change and the evolution of the availability of some resources, such as the memory, the battery level.

***Analyze.*** When a contextual change occurs, the cloud must analyze the relationship between this change and the dynamic variation points when configuring the requested services. So, the runtime environment must be monitored to detect contextual changes that might affect the cloud service configuration.

***Plan.*** The manager agent uses the consumer knowledge base and the cloud knowledge base fulfilled respectively by the consumer agents and the provider services agents. The cloud knowledge base contains cloud services reification and cloud features which may be, resources, constraints, etc.

The agents used in this work are proactive, intelligent and have goals. Therefore, we propose to use goals to select and represent requirements and the various objectives that the cloud service should realize. The approach uses the goal-oriented analysis model for modeling and reasoning about the variability model.

The connection between these two models is made using the mapping, as depicted in Figure 3, that connects each feature or requirement to a goal that defines the cloud environment. The agent manager has all the information about the commonalities and variability of a particular cloud environment, so it can define the cloud environment features model. Our approach is based especially on extended FM with cardinalities, attributes, and constraints. Indeed, the manager agent selects the required cloud services that meet the functional and non-functional requirements of the consumer demand, for example, the MySql database application for a request for cloud storage.

The features associated with the items selected in the goal model are selected in the different feature models if they are provided by the cloud environment. For example, the Mysql feature will be selected in all feature models. After the reconfiguration of the service, the agent manager keeps track of the configuration in the reusable database. Storing trace entries on reconfigurations in the context of the experiment provides information for a posteriori analysis, which ranges from context conditions to reconfiguration plans.

Once a cloud environment has been selected and configured, the cloud-related engine is launched to derive the configuration files and associated scripts, for example, a set of configuration commands to execute.

***Execute.*** Although the configuration files are required for proper configuration of the cloud environment, configuration scripts must be run by the agent manager to configure the environment.

*3.3 Metamodels and Mapping Rules*

Our approach utilizes an extended version of FM with cardinalities and attributes which permit the specification of constraints (Czarnecki, Helsen, & Eisenecker, 2004). This version offers an enhanced semantic expressiveness and it consequently enables more performance. Also, our approach adopts the feature model approach to express the variability of the cloud and the user environment. The cloud knowledge base and the consumer knowledge base were introduced as feature models, where features are elements of the base.

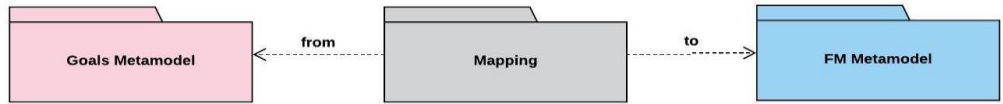

Figure 3. Metamodels Relationship

The manager agent is responsible for selectiong and configuring the features models of the cloud services using the mapping relationship between the goal model and feature model which must conform to their metamodels (Figure 4).

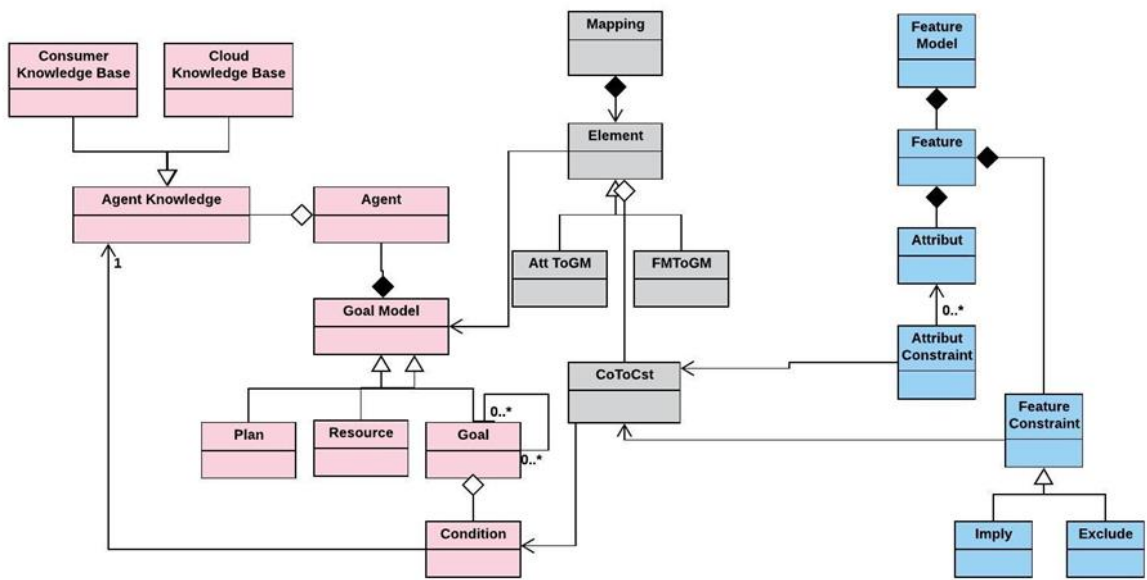

Figure 4. Approach metamodel

## 4. Case Study

We consider a motivating example for mobile cloud services for Diabetes Management. Diabetes is viewed, referring to the World Health Organization, as an incurable disease that affects 194 million people worldwide (Hamdan, 2011). It reaches epidemic proportions (Gan, 2003) in developing and newly industrialized nations. These high percentages are the result of the wrong management of diabetes and the poor quality of the existing offered services.

Diabetes needs an efficient and self-management service that is patient context-aware and interactive to every change in the patient data to keep it under continual and regular control. The service should provide the patient with his diabetes measurement, guide him in case of an increase or decrease of blood glucose level, save the patient records, reminds the patient about check-ups medication time and date.

We conduct a case study of applying the proposed approach to the domain of Mobile-Cloud Management service for diabetes with a Mobile Internet Device, which can provide the current location and information of the patient. The monitored context defines the current state of User(patient) Preference, the Device Context, and Situational Context. patient preference defines specific preference settings, particularly those have a relation with services selection and invocation. The Device Context represents the environmental settings and configurations of the patient's device.

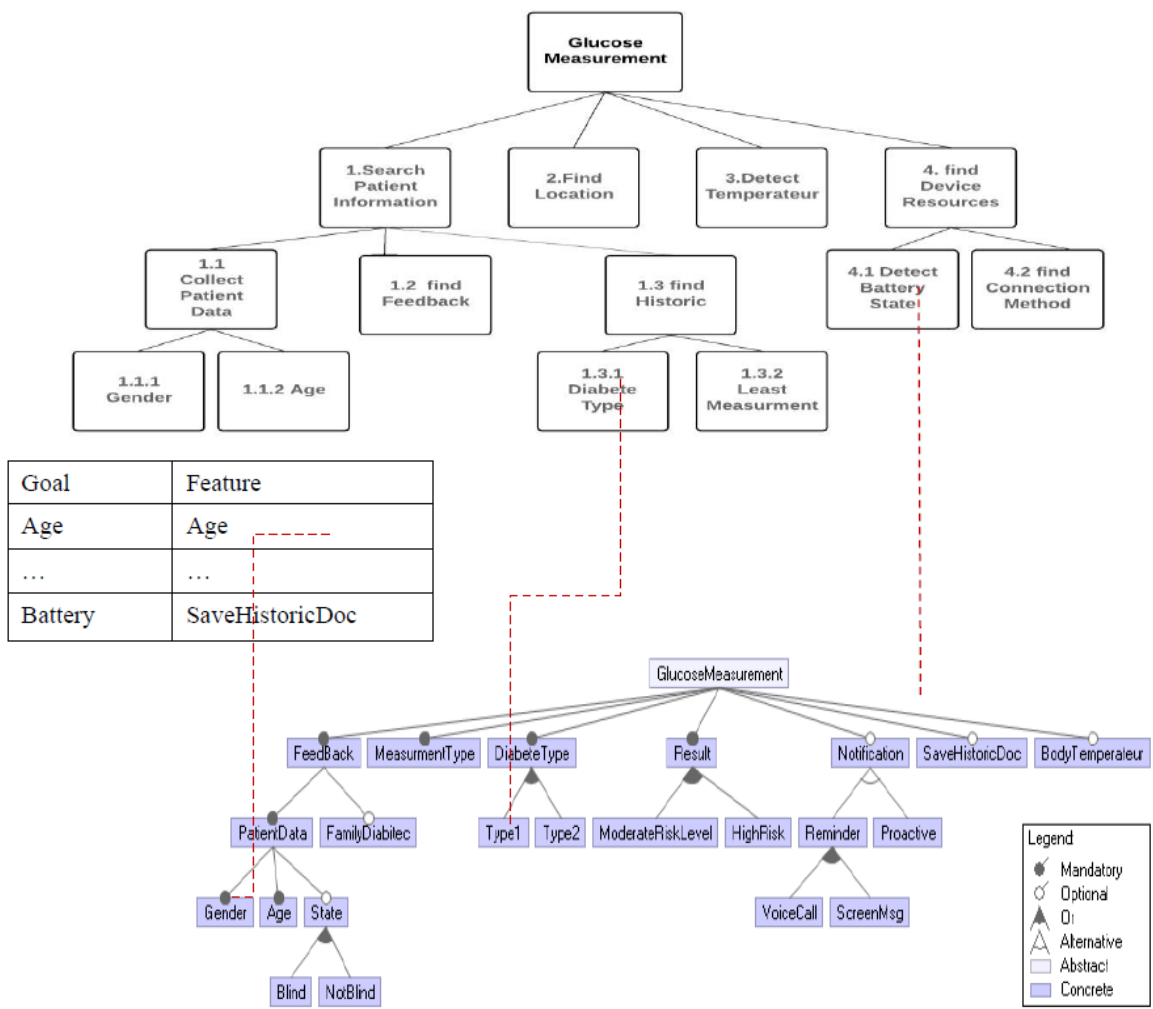

Figure 5. selection and configuration of the cloud service "Glucose Measurement"

According to the Situational Context, it contains a set of monitored data and information related to the user's location, level of Blood Glucose measurement, time, current lifestyle, weight, temperature, and others. The manager agent (Brocker agent) configures the "Glucose measurement" service basing on the goal model (Figure 3) provided by the consumer agent in the monitor phase of the MAPE-K mode. To achieve goals, agents are created Based on the goal model presented by the identification and specification of their plans. In the Plan phase, the goal model and Feature model are linked using a mapping basing on an Agent Definition Files (ADF) which is an XML File (Figure 4) that declares static agent properties.   In our example, to measure the glucose level of the patient, the manager agent needs a set of information from the consumer agent, like patient data, location, temperature, and other information about the device such as the device energy.  The potential of mobile offloading mainly depends on mobile network technologies such as WIFI and cellular.   However, data transmission using the WIFI network requires a considerable amount of energy from the mobile device as opposed to the cellular network. In the Analyze phase, when the battery level, for example, becomes low at run-time, the manager agent reconfigures the service by deselecting the feature 'SaveHistoricDoc' for not saving the historic documents on the cloud.

```xml
<agent xmlns="http://jadex.sourceforge.net/jadex-bdi"
       xmlns:xsi="http://www.w3.org/2001/XMLSchema-instance"
       xsi:schemaLocation="http://jadex.sourceforge.net/jadex-bdi
                           http://jadex.sourceforge.net/jadex-bdi-2.0.xsd"
  name="Buyer" package="jadex.bdi.examples.booktrading.buyer">

  <achievegoal name="find_Device_Resources">
  <value>$Low</value>
  <unique/>
  <creationcondition language="jcl">
    $beliefbase.eatingSaveHistoricDoc_NotAllowed
  </creationcondition>
  <dropcondition language="jcl">
    ......
  </dropcondition>
  <deliberation>
    <inhibits ref="wander_around"/>
  </deliberation>
</achievegoal>

</agent>
```

Figure 6. ADF file for "find_Device_Resources" agent

## 5. Related Work

We examine, in this section, different approaches that treat the deployment of applications in cloud environments. Then we compare, in table 2, these approaches to define their limits. The comparison is based on common characteristics shared by the studied approaches.

- **Approach A:** (Multi-Agent for monitoring cloud resources) (Alwada & Falah, 2018) presents an architecture based multi-agent system that is used by cloud computing to select the best resources and to create a negotiation method among the cloud consumers and the cloud providers.

- **Approach B:** (Infrastructure migration to the cloud) (Garc á-Gal án, Trinidad, Rana, & Ruiz-Cort és, 2016) addresses the problem of selecting the most suitable configuration among the configurations space provided by proposing an application based on FM. The approach was implemented and compared to two well-known business tools (Amazon TCO and CHloudScreener). However, for the moment, it can only handle the Amazon EC2 configuration space.

- **Approach C:** (selecting and configuring cloud environments) (Quinton, Romero, & Duchien, 2014) proposes SALOON, a software product lines-based platform to handle the variability of cloud providers and the resources at the configuration and deployment stage. They based on features models to represent this variability. This approach treats the variability only in the design phase and doesn't take into consideration the context changes at the runtime.

- **Approach D:** (Configuring Cloud Robotics Applications) (Gherardi, Hunziker, & Mohanarajah, 2014) explored the use of Extended Feature Models (EFMs) to provide support for the design and configuration of robotic applications when migrating to a cloud environment. The Models can be automatically transformed into XML and JSON configuration files offering support for the application migration to the cloud. However, they did not handle constraints when several robots share the same cloud components. Moreover, the feature selection was done with a manual manner resulting in a complex and error-prone task.

Table 2. A Comparative Study on Variability Management Approaches for Cloud Services Selection

| Approach | Approach A | Approach B | Approach C | Approach D | Our Approach |
|---|---|---|---|---|---|
| Cloud Application Area | IaaS | IaaS | PaaS, IaaS | SaaS, PaaS | SaaS, PaaS, IaaS |
| Phase of System Variability | Design time | Design time | Design time | Design time | Design and Runtime |
| Adaptation Type | Dynamic | Dynamic | Dynamic | Dynamic | Dynamic |
| Adaptation Space | Non functional | Non functional | Functional and non-functional | Functional and non-functional | Functional and non-functional |
| Flexibility | Not proposed | Multi-tenant Service | Not proposed | Not proposed | Use MAPL -K |
| Reusability | Not Proposed | Use SPLE | Use SPLE | Use SPLE | Use DSPL |
| Scalability | Use Multi-Agents System method | Not proposed | Automating the FM configuration files generation | Use PaaS | Use Multi-Agents System oriented Software |

## 6. Conclusion

In this work, the dynamic software product line approach and the multi-agent systems with the help of the MAPE-K reference model were introduced to provide support for dynamically selecting and configuring cloud environments for context-aware systems. This support takes into account the continuous change of consumer context and the system to adapt dynamically the reconfiguration of the cloud environment. Indeed, it reconfigures the cloud service at runtime to meet the consumer requirements that change in real-time (example insufficiency of a resource). The agent-oriented goal model and feature model are used to automatize the selection of the cloud service that meets the consumer requirements at the design and execution phase. The approach relies on the cloud and consumer knowledge model that contains all required data, attributes, and constraints on both environments provided by the agents. In our work, the use of a knowledge metamodel is motivated by the need to support the sharing of the knowledge base and reuse of knowledge by different consumers and even by other clouds. Mapping rules between the goal model and the feature model were adopted to allow an automated configuration of the different FMs. Indeed, the proposed approach permits an autonomy reconfiguration according to the context change. In progress, we intend to extend our approach to take into account the FM evolution with attributes and cardinalities. Indeed, the feature models with SPL evolve overtime (Sampaio, Borba, & Teixeira, 2019) to deal with the cloud market that evolves continuously. Consequently, it may occur inconsistencies between values defined and features cardinalities. In addition, we plane to employ our approach in other case studies to provide further evaluation of it.